\documentclass[journal, letter, 9pt]{IEEEtran}
\usepackage{amsmath}
\usepackage{amsfonts}
\usepackage{graphics}
\usepackage{array}
\usepackage{shortvrb}
\usepackage{epsf}
\usepackage{graphicx}
\usepackage{rotating}
\usepackage{multirow}
\usepackage{cite}
\usepackage{xcolor}
\usepackage{alltt}
\usepackage{soul}
\usepackage{overpic}
\usepackage{pict2e}
\usepackage{comment}
\usepackage{acronym}
\usepackage{float}
\usepackage[caption=false,font=footnotesize]{subfig}
\usepackage{placeins}
\usepackage{threeparttable}
\usepackage{booktabs}

\acrodef{cig}[CIG]{converter-interfaced generator}
\acrodef{coi}[COI]{Center of Inertia}
\acrodef{avr}[AVR]{Automatic Voltage Regulator}
\acrodef{agc}[AGC]{Automatic Generation Control}
\acrodef{lic}[LIC]{Local Integration Control}
\acrodef{pfc}[PFC]{Primary Frequency Control}
\acrodef{sm}[SM]{synchronous machine}
\acrodef{der}[DER]{Distributed Energy Resource}
\acrodef{ffr}[FFR]{Fast Frequency Regulation}
\acrodef{pll}[PLL]{Phased-Locked Loop}
\acrodef{rocof}[RoCoF]{Rate of Change of Frequency}
\acrodef{dae}[DAE]{Differential-Algebraic Equation}
\acrodef{pi}[PI]{Proportional-Integral}
\acrodef{pod}[POD]{Power Oscillation Damper}
\acrodef{pf}[PF]{Participation Factor}
\acrodef{lep}[LEP]{Linear Eigenvalue Problem}
\acrodef{ess}[ESS]{Energy Storage system}
\begin{document}
\title{Dual Grid-Forming Converter}
\newcommand{\T}{^{\scriptscriptstyle \rm T}}
\newcommand{\wcoi}{\omega_{\scriptscriptstyle \rm COI}}
\newcommand{\dwcoi}{\dot\omega_{\scriptscriptstyle \rm COI}}
\newcommand{\vd}{v_{d}}
\newcommand{\vq}{v_{q}}
\newcommand{\dvd}{\dot{v}_{d}}
\newcommand{\dvq}{\dot{v}_{q}}
\newcommand{\vdt}{v'_{d}}
\newcommand{\vqt}{v'_{q}}
\newcommand{\dvdt}{\dot{v}'_{d}}
\newcommand{\dvqt}{\dot{v}'_{q}}
\newcommand{\wt}{\tilde{\omega}}

\newcommand{\jj}{\jmath}

\author{
  Federico Milano, {\em IEEE Fellow} %
  \thanks{F.~Milano is with the School of Electrical
    and Electronic Engineering, University College Dublin (UCD), Belfield Campus, D04V1W8,
    Dublin, Ireland.  e-mail: \mbox{federico.milano@ucd.ie}}%
  %\vspace{-7mm}
}

\markboth{Submitted to IEEE PES Letters}{}
\maketitle
\begin{abstract}
  This letter proposes a \textit{dual} model for grid-forming~(GFM) controlled converters.  The model is inspired from the observation that the structures of the active and reactive power equations of lossy synchronous machine models are almost symmetrical in terms of armature resistance and transient reactance.  The proposed device is able to compensate grid power unbalance without requiring a frequency signal.  In fact, the active power control is based on the rate of change of the voltage magnitude.  On the other hand, synchronization and frequency control is obtained through the reactive power support. The letter shows that the proposed dual-GFM control is robust and capable of recovering a normal operating condition following large contingencies, such as load outages and three-phase faults.
\end{abstract}
\IEEEpeerreviewmaketitle

%\vspace{-1mm}

%======================================================================
\section{Introduction}
\label{sec:intro}

In recent years, a large variety of studies have appeared on the so-called grid-forming controlled converters (GFMs) \cite{rosso2021grid}.  The common understanding is that these devices are substantially resembling synchronous machines, the main difference being that one can tune their damping, which in a GFM control is not associated with friction but, rather, with a droop control \cite{rosso2021grid}; and the inertia constant, which, in the GFM control can be also set to zero, thus, \textit{de facto} avoiding the oscillatory behavior of synchronous machines \cite{Dhople24}.

Duality is another interesting aspect that has been discussed for GFM controlled converters.  In \cite{dualgfmgfl}, GFMs are considered the dual of grid-following converters (GFLs).  That is, the duality is between voltage source (GFM) and current source (GFL).
In this work, we propose an alternative approach, which, while taking inspiration from the synchronous machine model, constitutes a new type of duality.  

Based on the recently proposed concept of complex frequency \cite{freqcomplex}, the contributions of the letter are two-fold: (i) show that it is possible to define a grid-forming control strategy that is structurally different from synchronous machines; and (ii) show that the power balance of a power system can be maintained without having to rely on the measurement of the frequency.  This appears particularly relevant as converters do not necessarily link active power and frequency as synchronous machines do.

In particular, this letter shows how to set up a GFM control that utilizes the instantaneous bandwidth (voltage magnitude time derivative) rather than the instantaneous frequency (voltage phase angle time derivative) as slack variable to maintain the power balance of the grid.  The proposed device is also dual in terms of active and reactive power controllers.  The active power is utilised to keep null the voltage deviations and the reactive power to regulate frequency deviations.  The case studies show that the proposed dual-GFM control is stable and can maintain the power balance of the system equally well as synchronous machines and conventional GFM converters.

% ======================================================================
\section{Rationale and Proposed Dual GFM Control Model} 
\label{sec:theory}
% ======================================================================

\begin{figure}[ht!]
  \begin{center}
    \resizebox{0.75\linewidth}{!}{\includegraphics[scale=1.0]{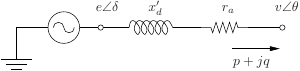}}
    \caption{Classical transient model of the synchronous machine with
      inclusion of armature losses.}
    \label{fig:machine}
  \end{center}
  \vspace{-3mm}
\end{figure}

Consider the power injections of the lossy electromechanical model of the synchronous machine shown in Fig.~\ref{fig:machine}:
\begin{align}
  \label{eq:p}
  p &= \frac{ [ev \cos(\delta-\theta) - v^2] r_a +
      [ev\sin(\delta -\theta)] x'_d }{r_a^2 + {x'_d}^2} \, ,  \\
  \label{eq:q}
  q &= \frac{ [ev \cos(\delta-\theta) - v^2] x'_d -
      [ev\sin(\delta -\theta)] r_a}{r_a^2 + {x'_d}^2} \, .
\end{align}

In synchronous machines, the armature resistance $r_a$ is small with respect to $x'_d$ and is often neglected, thus leading to the well-known equations:
\begin{align}
  \label{eq:psyn}
  p &= \frac{ev\sin(\delta -\theta)}{x'_d} \, , \\
  \label{eq:qsyn}
  q &= \frac{ev \cos(\delta-\theta) - v^2}{x'_d} \, .
\end{align}
This work, on the other hand, considers the \textit{dual} parts of \eqref{eq:p} and \eqref{eq:q}, that is, the terms that depend on the armature resistance and suppose that the reactance $x'_d$ is zero or negligible.  This can be done because a converter is not a machine with physical coils and its parameters can be tuned as desired.  This leads to:
\begin{align}
  \label{eq:pdual}
  \tilde{p} &= \frac{ev\cos(\delta -\theta) - v^2}{r_a} \, , \\
  \label{eq:qdual}
  \tilde{q} &= -\frac{ev \sin(\delta-\theta)}{r_a} \, .
\end{align}
Finally, assume that the virtual parameter that represents the armature resistance is negative, say $K = -1/r_a$, thus leading to:
\begin{align}
  \label{eq:pgfm}
  \tilde{p} &= K v^2 - K ev\cos(\delta -\theta) \, , \\
  \label{eq:qgfm}
  \tilde{q} &= K ev \sin(\delta-\theta) \, .
\end{align}
Equations \eqref{eq:pgfm} and \eqref{eq:qgfm} describe a device where the active power strongly depends on the magnitude of the internal emf $e$, while the reactive power strongly depends on the phase angle $\delta$.  

\subsection{Dual Swing Equations}

The proposed model is completed with a dual swing equation, as follows.  First, recall that the conventional swing equation is defined in terms of the machine rotor angle:
\begin{equation}
  \label{eq:swing}
  \begin{aligned}
    \dot \delta &= \omega -\omega_o \, , \\
    M \dot{\omega} &= p_m - p(e, v, \delta, \theta) - D (\omega -\omega_o)  \, ,
  \end{aligned}
\end{equation}
where $p$ is given in \eqref{eq:psyn}, $\omega$ is the angular rotor speed with respect to the synchronous reference angular frequency $\omega_o$; $p_m$ is the mechanical power as imposed by the machine turbine governor; $M$ is the mechanical starting time; and $D$ is the damping coefficient.

To obtain the dual swing equation, consider the complex quantity:
\begin{equation}
  \label{eq:e1}
  \bar{e} = e \, {\rm exp}(j \, \delta) \, .
\end{equation}
where $e$ and $\delta$ are assumed to be time varying.  Then, defining $u = \ln(e)$, $e \ne 0$, \eqref{eq:e1} becomes:
\begin{equation}
  \label{eq:e2}
  \bar{e} = {\rm exp}(u + j \, \delta) \, ,
\end{equation}
and its time derivative is:
\begin{equation}
  \label{eq:edot}
  \dot{\bar{e}}
  = (\dot{u} + j \, \dot{\delta}) \, {\rm exp}(u + j \, \delta)
  = (\varrho + j \, \omega) \, \bar{e} \, ,
\end{equation}
where $\omega$ is defined as in \eqref{eq:swing} and $\varrho$ is:
\begin{equation}
  \label{eq:rho}
  \varrho = \dot{u} = \dot{e} / e \, .
\end{equation}
Finally, the swing equation dual to \eqref{eq:swing} is defined as:
\begin{equation}
  \label{eq:dual0}
  \begin{aligned}
  \dot{u} &= \rho \, , \\
  \tilde{M} \dot{\rho} &= p^{\rm ref} - \tilde{p}(u, v, \delta, \theta) - \tilde{D} \rho  \, ,
  \end{aligned}
\end{equation}
or, equivalently
\begin{equation}
  \label{eq:dual}
  \begin{aligned}
    \dot{e} &= \varrho \, e \, , \\
    \tilde{M} \dot{\varrho}
            &= p^{\rm ref} -
              \tilde{p}(e, v, \delta, \theta) -
              \tilde{D} \varrho \, , 
  \end{aligned}
\end{equation}
where $\tilde{M}$ and $\tilde{D}$ are parameters that resemble a virtual inertia and a virtual damping, respectively; $p^{\rm ref}$ is the active power as defined by the control of the converter; and $\tilde{p}$ is given in \eqref{eq:pgfm}.  As $\omega$ in \eqref{eq:swing}, also $\varrho$ in \eqref{eq:dual} is referred to a reference value, say $\varrho_o$.  However, $\varrho_o \equiv 0$ as the bus voltage magnitudes are required to be constant in steady state. 

\subsection{Dual Primary Controllers}

The primary active power control for the synchronous machine is the turbine governor which tracks, typically using a drop, the machine's rotor speed.  The simplest first order model for the turbine governor can be written as:
\begin{equation}
  \label{eq:wcontrol}
  \begin{aligned}
    T_m \dot{p}_m &= \frac{1}{R} (\omega^{\rm ref} - \omega) + p_{m,o} - p_m \, ,
  \end{aligned}
\end{equation}
where $R$ is the droop coefficient and $p_{m, o}$ is the power set point of the turbine.  In the same vein, the primary active power control for the dual-GFM control is required to track $\varrho$, for example:
\begin{equation}
  \label{eq:rcontrol}
  \begin{aligned}
    \tilde{T}_m \dot{p}^{\rm ref}
    &= \frac{1}{\tilde{R}}
      (\varrho^{\rm ref} - \varrho) + p^{\rm ref}_o - p^{\rm ref} \, ,
  \end{aligned}
\end{equation}
where $\varrho^{\rm ref} = \varrho_o = 0$ and $p^{\rm ref}_o$ is the converter power set point.  As $\varrho$ must be zero in steady state, to avoid a drift of $e$, the active power primary control of the dual-GFM control can be perfect tracking ($\tilde{R} \rightarrow 0$).

The reactive power control for the synchronous machine is obtained by regulating the field voltage through an exciter.  For a 3-rd order machine model, a basic automatic voltage control has the form:
\begin{equation}
  \label{eq:vcontrol}
  \begin{aligned}
    T'_{d0} \dot{e} &=  v_f - (x_d - x'_d) i_d - e \, , \\
    T_r \dot{v}_f &= K_r (v^{\rm ref} - v) - v_f \, ,
  \end{aligned}
\end{equation}
where $v_f$ is the field voltage, $i_d$ is the $d$-axis machine stator current and $e \equiv e'_q$ is the internal $q$-axis transient voltage of the machine; and the parameters have the usual meaning.

For the dual-GFM control, the reactive power can be regulated directly through the angle $\delta$.  The dual to \eqref{eq:vcontrol} equation is thus:
\begin{equation}
  \label{eq:dcontrol}
  \begin{aligned}
    T_q \dot{\delta} &=  K_q (q^{\rm ref} - \tilde{q}) - \delta \, , \\
    \tilde{T}_r \dot{q}^{\rm ref} &= \tilde{K}_r (\omega^{\rm ref} - \omega) - q^{\rm ref} \, ,
  \end{aligned}
\end{equation}
where the expression of $\tilde{q}$ is given in \eqref{eq:qgfm}.  The similarity between \eqref{eq:vcontrol} and \eqref{eq:dcontrol} can be better appreciated by defining:
\begin{equation}
  \delta_r = K_q \, q^{\rm ref} \, ,
\end{equation}
which leads to rewrite \eqref{eq:dcontrol} as:
\begin{equation}
  \label{eq:dcontrol2}
  \begin{aligned}
    T_q \dot{\delta} &=  \delta_r - K_q \, \tilde{q} - \delta \, , \\
    \tilde{T}_r \dot{\delta}_r &= \tilde{K}'_{r} (\omega^{\rm ref} - \omega) - \delta_r \, ,
  \end{aligned}
\end{equation}
where $\tilde{K}'_{r} = \tilde{K}_r / K_q$.

\begin{figure}[ht!]
  \begin{center}
    \resizebox{0.99\linewidth}{!}{\includegraphics[scale=1.0]{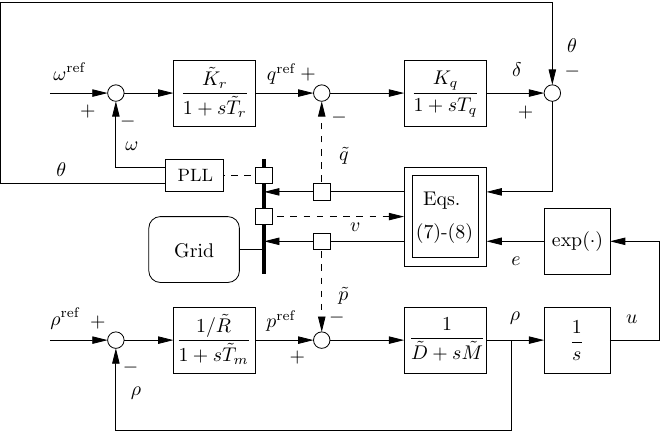}}
    \caption{Control scheme of the dual-GFM converter.}
    \label{fig:dualgfm}
  \end{center}
  \vspace{-3mm}
\end{figure}

The resulting dual-GFM control scheme is shown in Fig.~\ref{fig:dualgfm}.  Note that the bus voltage phase angle and frequency are assumed to te obtained using a standard PLL.  PLL-free solutions --- as in conventional GFM converters --- can be adopted, but this implementation aspect is beyond the scope of the letter.

It is important to note that the virtual angular speed does not appear in the formulation of the dual-GFM control except for the time derivative of the internal signal $\delta$ in \eqref{eq:dcontrol}.  This is consistent with synchronous machines, where $\varrho$ does not appear anywhere except for the time derivative of the internal emf $e$ in \eqref{eq:vcontrol}.

It is still necessary, of course, to fix the frequency in the system.  As shown in Fig.~\ref{fig:dualgfm}, the dual-GFM converter imposes the frequency at the bus through the regulation of the reactive power.  This makes sure that there is no drift in the phase angles of the voltages of the system and, hence, the frequency of $e$ is $\omega^{\rm ref}$ in steady-state.  As $\delta$ is relative to the phase angle $\theta$ of the voltage at the point of connection of the converter through \eqref{eq:pgfm} and \eqref{eq:qgfm}, in steady state, also the rest of the grid is synchronous at the rated frequency $\omega^{\rm ref}$.  In this way, the dual-GFM control is also able provide a perfect tracking (or almost perfect tracking as the one illustrated in Fig.~\ref{fig:dualgfm}) control of the frequency without the need of a secondary frequency control as it happens for synchronous machines and conventional GFM converters.

Finally, note that the objective of the letter is not to show that $p$-$v$ and $q$-$\omega$ controllers can be effective, as it has been done in some recent works, e.g., \cite{weilin, RuiLi:2024}, but, rather, to show that a power system can be operated exclusively with GFM converters that are structurally different from synchronous machines and conventional GFM converters.

\vspace{-2mm}

% ======================================================================
\section{Case Study}
\label{sec:cstudy}
% ======================================================================

In this section, we illustrate the performance of the proposed control with the WSCC 9-bus system and a 1479-bus dynamic model of the all-Ireland transmission system.  Simulation results are obtained with the software tool Dome \cite{dome}.

\vspace{-2mm}

\subsection{WSCC 9-bus System}

\begin{figure}[ht!]
  \centering
  \resizebox{0.95\linewidth}{!}{\includegraphics{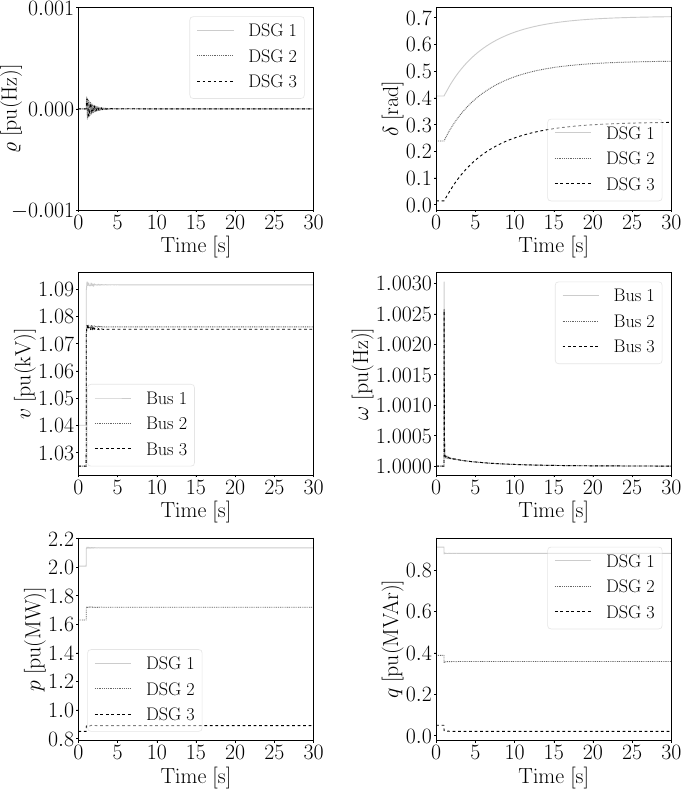}}
  \caption{WSCC 9-bus system -- Behavior of relevant variables of the dual-GFMs connected at buses 1-3 following a $20\%$ loss of load at bus 5.}
  \label{fig:wscc1}
  \vspace{-3mm}
\end{figure}

\begin{figure}[ht!]
  \centering
  \resizebox{0.95\linewidth}{!}{\includegraphics{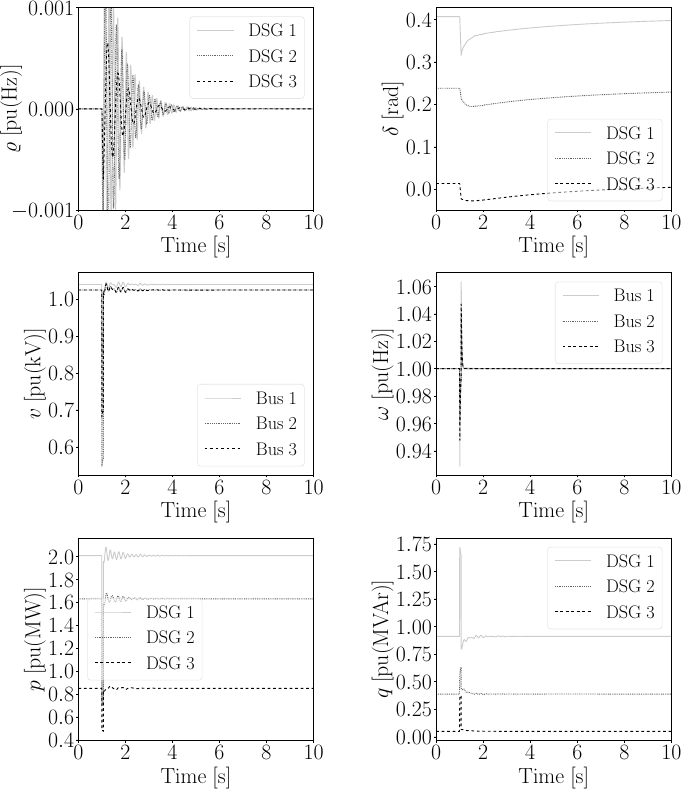}}
  \caption{WSCC 9-bus system -- Behavior of relevant variables of the dual-GFMs connected at buses 1-3 following a fault at bus 7 occurring at $t=1$ and cleared after $60$ ms.}
  \label{fig:wscc2}
  \vspace{-3mm}
\end{figure}

The original WSCC 9-bus system is modified by removing the 3 synchronous machines and replacing them with dual-GFM devices and the primary controllers defined in the previous section.  PSS are also included and properly tuned to damp the oscillations of the dual-GFMs. Figures \ref{fig:wscc1} and \ref{fig:wscc2} shows the performance of relevant quantities of the dual-GFMs following, respectively, a load outage and a three-phase fault cleared after three cycles.  In both cases, the dual-GFMs are capable of restoring normal operating conditions.  As there is no synchronous machine or conventional GFM converter, no device maintains the power balance relying on a measurement of the instantaneous frequency.  The following parameters for the dual-GFM are used: $K=0.1$, $\tilde{M} = 30$ s, $\tilde{D} = 20$, $\tilde{T}_m = 2$ s, $\tilde{R} = 0.05$, $K_q = 10$, $T_q = 5$ s, $\tilde{K}_r = 40$, $\tilde{T}_r = 1$ s.  $K$, $\tilde{M}$, $\tilde{D}$ and $\tilde{R}$ are in pu w.r.t. the capacity of the converter.

\subsection{All-Island Irish Transmission System}

This section illustrates the dynamic performance of the proposed dual-GFM converter for a dynamic model of the all-island Irish transmission system.  As starting point, we have utilized a dynamic model of the all-island, Irish power system that includes 1479 buses, 1851 transmission lines and transformers, 22 synchronous generators, along with their appropriate control systems, 169 wind power plants and 245 loads.  All wind power plants are assumed to be GFLs and not to provide any inertial response nor fast-frequency regulation.  More details on this model can be found in \cite{tidalgen}.  In this study, all synchronous machines are substituted with the proposed dual-GFM using same parameters as in the previous section except for $K = 1$, $\tilde{M} = 15$ s, $\tilde{D} = 0.5$.  This leads to a system where there is no conventional inertial response nor frequency control and where the only conventional voltage support, i.e., based on reactive power, is provided by the voltage regulators of the wind power plants.

\begin{figure}[ht!]
  \centering
  \resizebox{0.95\linewidth}{!}{\includegraphics{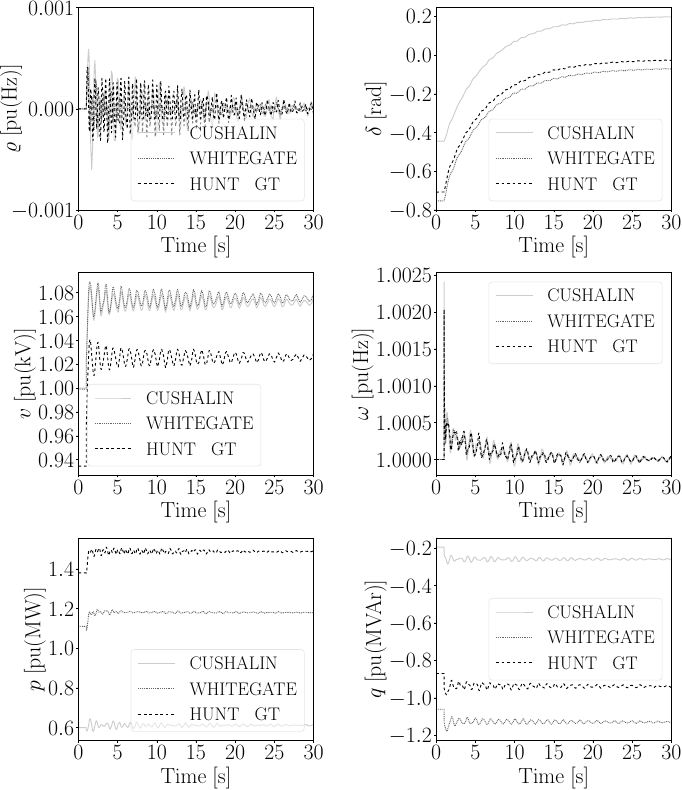}}
  \caption{All-island Irish transmission system -- Behavior of relevant variables of three dual-GFMs following the outage of a large load.}
  \label{fig:eire}
  \vspace{-3mm}
\end{figure}

Figure \ref{fig:eire} shows the dynamic performance of the modified model of the Irish transmission system following the outage of a large load.  As for the example for the WSCC 9-bus system, the dual-GFMs cope well with the power imbalance and recover an normal operating condition.

% ======================================================================
\section{Conclusions}
\label{sec:Conclusion}
% ======================================================================

The letter describes a novel approach for GFM controlled converters.  The proposed model is based on a duality between instantaneous bandwidth (i.e., rate of change of the voltage magnitude) and instantaneous frequency (i.e., rate of change of the voltage phase angle) of the virtual internal emf of GFM converters.  The resulting device is able to keep the system's balance following large disturbances such as loss of loads and faults, without the need for frequency measurements.  In this dual model, the instantaneous bandwidth takes the same role as the instantaneous frequency in synchronous machines and the synchronization is obtained through the reactive power control.

Simulation results indicate that the dual-GFM appears to be particularly robust and stable following large contingencies.  Moreover, as it utilizes the reactive power regulate frequency, the dual-GFM approach might be more effective than conventional GFM converters in distribution and low voltage networks.  Finally, the lack of dependency on the frequency for maintaining the power balance of the grid also suggests that the proposed model can be utilized in dc systems.

Future work will further investigate the possibilities that the proposed complex-frequency duality offers in terms of the definition of novel GFM converter controllers \color{black} as well as its potential application to dc grids.

\newpage


% ======================================================================

\vfill


\begin{thebibliography}{1}
\providecommand{\url}[1]{#1}
\csname url@samestyle\endcsname
\providecommand{\newblock}{\relax}
\providecommand{\bibinfo}[2]{#2}
\providecommand{\BIBentrySTDinterwordspacing}{\spaceskip=0pt\relax}
\providecommand{\BIBentryALTinterwordstretchfactor}{4}
\providecommand{\BIBentryALTinterwordspacing}{\spaceskip=\fontdimen2\font plus
\BIBentryALTinterwordstretchfactor\fontdimen3\font minus
  \fontdimen4\font\relax}
\providecommand{\BIBforeignlanguage}[2]{{%
\expandafter\ifx\csname l@#1\endcsname\relax
\typeout{** WARNING: IEEEtran.bst: No hyphenation pattern has been}%
\typeout{** loaded for the language `#1'. Using the pattern for}%
\typeout{** the default language instead.}%
\else
\language=\csname l@#1\endcsname
\fi
#2}}
\providecommand{\BIBdecl}{\relax}
\BIBdecl

\bibitem{rosso2021grid} R.~Rosso, X.~Wang, M.~Liserre, X.~Lu, and S.~Engelken, ``Grid-forming converters: Control approaches, grid-synchronization, and future trends—a review,'' \emph{IEEE Open Journal of Industry Applications}, vol.~2, pp.  93--109, 2021.

\bibitem{Dhople24} M.~Lu, W.~Cai, S.~Dhople, and B.~Johnson, ``Large-signal stability of phase-balanced equilibria in single-phase grid-forming inverter systems,'' \emph{IEEE Transactions on Power Electronics}, vol.~39, no.~3, pp. 3623--3636, 2024.

\bibitem{dualgfmgfl} Y.~Li, Y.~Gu, and T.~C. Green, ``Revisiting grid-forming and grid-following inverters: A duality theory,'' \emph{IEEE Transactions on Power Systems}, vol.~37, no.~6, pp. 4541--4554, 2022.

\bibitem{freqcomplex} F.~Milano, ``Complex frequency,'' \emph{IEEE Transactions on Power Systems}, vol.~37, no.~2, pp. 1230--1240, 2022.

\bibitem{weilin} W.~Zhong, G.~Tzounas, and F.~Milano, ``Improving the power system dynamic response through a combined voltage-frequency control of distributed energy resources,'' \emph{IEEE Transactions on Power Systems}, vol.~37, no.~6, pp.~4375-4384, November 2022.

\bibitem{RuiLi:2024} L.~Yu, Z.~Fu, R.~Li, and J.~Zhu, ``{DRU-HVDC} for offshore wind power transmission: {A} review,'' \emph{IET Renewable Power Generation}, vol.~18, no.~13, pp. 2080--2101, 2024.

\bibitem{dome} F.~Milano, ``A {Python}-based software tool for power system analysis,'' in \emph{IEEE PES General Meeting}, 2013, pp. 1--5.

\bibitem{tidalgen} G.~M. J{\'o}nsd{\'o}ttir and F.~Milano, ``Stochastic modeling of tidal generation for transient stability analysis: A case study based on the all-island irish transmission system,'' \emph{Electric Power Systems Research}, vol. 189, p. 106673, 2020.

\end{thebibliography}
\end{document}